%% file: main.tex
\documentclass{article}

\usepackage{microtype}
\usepackage{graphicx}
\usepackage{subcaption}
\usepackage{booktabs}

\usepackage{hyperref}

\usepackage[accepted]{icml2026}

\usepackage{amsmath}
\usepackage{amssymb}
\usepackage{mathtools}
\usepackage{amsthm}

\usepackage[capitalize,noabbrev]{cleveref}

\theoremstyle{plain}

\theoremstyle{definition}

\theoremstyle{remark}

\usepackage[textsize=tiny]{todonotes}

\usepackage[utf8]{inputenc}
\usepackage[T1]{fontenc}
\usepackage{hyperref}
\usepackage{url}
\usepackage{booktabs}
\usepackage{amsfonts}
\usepackage{nicefrac}
\usepackage{microtype}
\usepackage{xcolor}
\usepackage{graphicx}    
\usepackage{hyperref}    
\usepackage{tipa}        
\usepackage{multirow}
\usepackage{array}
\usepackage{subcaption}
\usepackage{caption}
\usepackage{kotex}

\icmltitlerunning{Beyond Binary Instrument QA: Probing Instrument Grounding in Music Audio-Language Models}

\begin{document}

\twocolumn[
  \icmltitle{Beyond Binary Instrument QA: Probing Instrument Grounding \\in Music Audio-Language Models}
  
  \icmlsetsymbol{equal}{*}

  \begin{icmlauthorlist}
    \icmlauthor{Yujun Lee}{skku}
    \icmlauthor{Joonhyeok Shin}{skku}
    \icmlauthor{Hyoeun Kim}{skku}
    \icmlauthor{Kyuhong Shim}{skku}
  \end{icmlauthorlist}

  \icmlaffiliation{skku}{Sungkyunkwan University}

  \icmlcorrespondingauthor{Yujun Lee}{yj090744@g.skku.edu}
  \icmlcorrespondingauthor{Kyuhong Shim}{khshim@skku.edu}

  \icmlkeywords{Instrument Grounding, Music Audio-Language Models, Question Anwering}

  \vskip0.3in
]

\printAffiliationsAndNotice{}

\begin{abstract}
Recent music audio-language models achieve high accuracy on instrument question-answering benchmarks, but it remains unclear whether this reflects robust audio grounding or benchmark-specific shortcuts. 
In this paper, we introduce an OpenMIC-derived diagnostic benchmark sequence for instrument grounding in music audio-language models, extending binary instrument-presence QA to genre-prior-reduced examples, confusable instrument discrimination, longer audio context, and temporal localization. 
Across these settings, high binary QA accuracy often fails to predict model behavior: models can exhibit option-position bias, confusable-instrument errors, and temporal response bias.
These results suggest that instrument grounding should be evaluated with multi-axis diagnostic benchmarks rather than a single aggregate accuracy.
\end{abstract}

\input{sections/sec01_intro}
\input{sections/sec03_method}
\input{sections/sec04_experiments}
\input{sections/sec05_conclusion}

\newpage
{
    \bibliography{references}
}

\bibliographystyle{icml2026}

\newpage
\appendix
\onecolumn
\input{sections/sec10_appendix}

\end{document}

%% file: sections/sec01_intro.tex
\section{Introduction}\label{sec:intro}

Recent audio-language models have rapidly expanded from general audio understanding to music-specific reasoning, building on a growing line of audio-language models~\cite{chu2023qwen, chu2024qwen2, kong2024audio, ghosh2025audio, tang2024salmonn, gong2024listen, gong2023joint}. 
Recent systems further extend this trend: Qwen2.5-Omni~\cite{xu2025qwen3} processes audio as part of multimodal interaction, Audio Flamingo 3  (AF3)~\cite{ghosh2026audio} targets broad audio understanding across speech, sound, and music, and Music Flamingo (MF)~\cite{ghosh2025music} specializes in music understanding tasks such as instrumentation, timbre, harmony, structure, lyrics, and temporal reasoning. 
As these models become more capable, how to evaluate musical understanding and interpret benchmark results becomes increasingly important~\cite{choi2017tutorial}.

\begin{table*}[ht!]
\centering
\caption{Overview of the diagnostic benchmark sequence. Binary QA (9,332) and genre-prior-reduced QA (590) report overall, Yes-QA, and No-QA accuracy. Instrument discrimination task (1,051) reports multiple-choice accuracy. Long-context multi-label benchmark (1,028) reports exact-set accuracy and F1. Temporal localization task (3,579) reports accuracy over three candidate time ranges.}
\label{tab:benchmark_sequence}
\resizebox{\textwidth}{!}{
\begin{tabular}{lcccccccccc}
\toprule
\multirow{2}{*}{Model}
& \multicolumn{3}{c}{Binary QA}
& \multicolumn{3}{c}{Prior-Reduced QA}
& {\shortstack{Discrimination}}
& \multicolumn{2}{c}{\shortstack{Long-Context}}
& {\shortstack{Temp. Loc.}} \\
\cmidrule(lr){2-4}
\cmidrule(lr){5-7}
\cmidrule(lr){8-8}
\cmidrule(lr){9-10}
\cmidrule(lr){11-11}
& Overall & Yes & No
& Overall & Yes & No
& Acc.
& Exact & F1
& Acc. \\
\midrule
MF
& 87.60 & 86.84 & 88.36
& 85.76 & 78.66 & 91.10
& 44.43
& 32.39 & 74.79
& 52.14 \\

MF-Think
& 81.75 & 68.05 & 95.46
& 81.02 & 60.87 & 96.14
& 47.76
& 35.80 & 72.72
& 44.90 \\

Qwen2.5-Omni
& 88.74 & 88.00 & 89.48
& 87.97 & 85.77 & 89.61
& 86.11
& 56.61 & 84.67
& 72.90 \\

AF3
& 87.19 & 81.29 & 93.10
& 84.75 & 74.31 & 92.58
& 68.41
& 24.03 & 74.80
& 33.70 \\

GPT-4o-audio
& -- & -- & --
& -- & -- & --
& 87.73
& 54.18 & 82.70
& 57.50 \\

Gemini 2.5 Pro
& -- & -- & --
& -- & -- & --
& 83.82
& 51.26 & 80.36
& 86.28 \\

Gemini 2.5 Flash
& -- & -- & --
& -- & -- & --
& 83.92
& 44.94 & 70.57
& 70.30 \\
\bottomrule
\end{tabular}
}
\end{table*}

Prior audio-language and music-language evaluation has used datasets and benchmarks for audio captioning, event recognition, and music understanding~\cite{gemmeke2017audio, kim2019audiocaps, drossos2020clotho, elizalde2023clap, doh2023lp, agostinelli2023musiclm, weck2024muchomusic, zhao2024openmu}.
For instrument-level evaluation, a common benchmark format is instrument-presence question-answering (QA), where a model answers whether a target instrument is present.

However, high accuracy in binary QA does not necessarily imply grounded instrument understanding.
A model can answer correctly by exploiting genre-instrument priors, response-format shortcuts, or short context cues, without reliably distinguishing instruments from the audio itself~\cite{geirhos2020shortcut, gururangan2018annotation, mccoy2019right}.
The limitation becomes more important when evaluation moves beyond isolated yes/no questions; robust instrument grounding should require a model to distinguish acoustically aor musically confusable instruments, recognize multiple instruments in longer mixtures, and localize when a target instrument appears.

In this paper, we investigate whether binary instrument-QA performance remains reliable under more diagnostic evaluation formats.
Using OpenMIC-2018~\cite{humphrey2018openmic}, we construct an instrument-grounding benchmark sequence, which starts from binary QA and progressively introduces genre-prior-reduced presence QA, confusion-aware instrument discrimination, multi-label recognition, and temporal instrument localization.

Our contributions are threefold. 
First, we introduce an OpenMIC-derived diagnostic benchmark sequence for probing instrument grounding in music audio-language models. 
Second, we evaluate recent general-purpose and music-specialized models and show that similar binary-QA accuracy can conceal substantially different failure modes.
Third, we analyze model behavior beyond aggregate accuracy, revealing option-position biases, instrument-label preferences, and temporal-range response biases that are not visible in standard instrument-presence binary QA.
We will release the benchmark metadata, prompt templates, and evaluation code to support reproducible comparison.

%% file: sections/sec03_method.tex
\section{Benchmark Construction and Evaluation}
\label{sec:benchmark_eval}

We construct an OpenMIC-derived diagnostic sequence~\cite{humphrey2018openmic} using the relevance annotations described in Appendix~\ref{app:openmic_format}: binary instrument-presence QA, genre-prior-reduced presence QA, confusion-aware instrument discrimination, long-context multi-label recognition, and temporal instrument localization.
This design follows the broader view that benchmark accuracy alone can miss systematic failure modes, motivating targeted diagnostic and behavioral tests~\cite{ribeiro2020beyond, geirhos2020shortcut, srivastava2023beyond}.

\subsection{Binary Instrument-Presence QA}
\label{sec:binary_qa}

We begin with binary instrument-presence QA.
Each example consists of a 10-second OpenMIC clip and a target instrument, and the model is asked whether the instrument is present.
From 4,666 clips, we generate one positive and one negative QA pair per clip, producing 9,332 examples (Appendix~\ref{app:binary_qa}).
As shown in Table~\ref{tab:benchmark_sequence}, this setting yields high accuracy across all evaluated models: MF, AF3, and Qwen2.5-Omni exceed 87\%, and MF-Think~\footnote{MF-Think denotes the same Music Flamingo model evaluated with reasoning-enabled inference.} reaches 81.75\%. 
The results indicate that binary QA is a relatively permissive evaluation format for current audio-language models.

This observation raises a diagnostic question: \textit{are models truly identifying the target instrument from the audio}, or are they leveraging genre-instrument associations and the simplicity of yes/no responses? 
We therefore next reduce the influence of genre-level priors.

\subsection{Genre-Prior-Reduced Presence QA}
\label{sec:genre_hard}

To examine the influence of genre-instrument priors, we construct a hard set from the 9,108 binary QA examples with usable genre metadata.
The examples are split into 70\% training and 30\% test data, stratified by the gold yes/no label.
A simple genre-prior baseline estimates the positive-answer rate for each genre-instrument pair from the training split, using an instrument-level fallback for unseen pairs.
Test examples incorrectly answered by this baseline are selected as hard cases, resulting in 590 examples (see Appendix~\ref{app:genre_hard}).

Overall accuracy decreases only moderately on this hard set.
This subset does not eliminate all genre-related cues, but reduces examples that are solved by a simple genre-instrument prior.
Nevertheless, reducing such genre-favored cases does not fully expose model limitations, since the task remains binary~\cite{gardner2020evaluating}.
We therefore remove the yes/no response format and require discrimination between confusable instruments.

\subsection{Confusion-Aware Instrument Discrimination}
\label{sec:confusion_mc}

The third benchmark changes the task from binary presence detection to two-way instrument discrimination.
Each example contains a 10-second music clip and two candidate instruments sampled from a predefined confusable instrument group(see Appendix~\ref{app:confusable_groups}), producing 1,051 multiple-choice examples (see Appendix~\ref{app:confusion_mc}).
Note that the confusable groups are not part of the original OpenMIC annotations, but are manually added during benchmark construction to define musically or acoustically related candidate sets.
Thus, the task should be interpreted as related-instrument discrimination rather than as a perceptually validated human-confusion benchmark.

This multiple-choice format evaluates whether models can distinguish between related candidate instruments rather than merely answer whether a named instrument is plausible.
This also enables response-format analysis by varying the answer interface, such as A/B labels, X/Y labels, or direct instrument-name output. 
As shown in Section 3, this benchmark reveals substantial performance gaps and response biases that are hidden by binary QA.
Since the input is still a single 10-second clip, we next extend the setting to longer 30-second concatenated music examples.

\subsection{Long-Context Multi-label Instrument Recognition}
\label{sec:strict30}

The fourth benchmark evaluates multi-label instrument recognition in a longer music context.
Multi-label evaluation is commonly used when multiple target classes can be simultaneously present, requiring metrics beyond single-label accuracy~\cite{zhang2013review}.
Each example is a 30-second music input generated by concatenating three 10-second OpenMIC clips.
Four candidate instruments are sampled from the same confusable group, two are present and the other two are absent.
The model must select all candidate instruments that appear in the 30-second input, producing 1,028 multi-label examples (see Appendix~\ref{app:strict30}).
 
The results show a clear gap between exact-set accuracy and partial recognition performance.
Exact-set accuracy ranges from 24.03\% to 56.61\%, while F1 is much higher at 70.57\%--84.67\%.
This indicates that models often recover part of the correct set but struggle to identify all present instruments exactly.
Since the task still asks only whether instruments appear somewhere in the input, the final benchmark introduces explicit temporal localization.

\subsection{Temporal Instrument Localization}
\label{sec:temporal_eval}

The final benchmark evaluates whether models can localize an instrument in time.
Unlike the preceding benchmarks, this task requires temporal grounding rather than only presence detection.
Each example is a 30-second input constructed by concatenating three 10-second OpenMIC clips.
For a target instrument, exactly one segment has a high-confidence positive label, while the other two have negative labels.
The model must choose the time range in which the target instrument appears.
This benchmark includes 3,579 examples, with 1,190 labeled as 0--10 sec, 1,195 as 10--20 sec, and 1,194 as 20--30 sec (see Appendix~\ref{app:temporal}).
Because the three time ranges are nearly balanced, a model cannot obtain high accuracy by exploiting a majority class.

%% file: sections/sec04_experiments.tex
\section{Analysis}\label{sec:analysis}

\begin{figure*}[!t]
\centering
\includegraphics[width=1.0\textwidth]{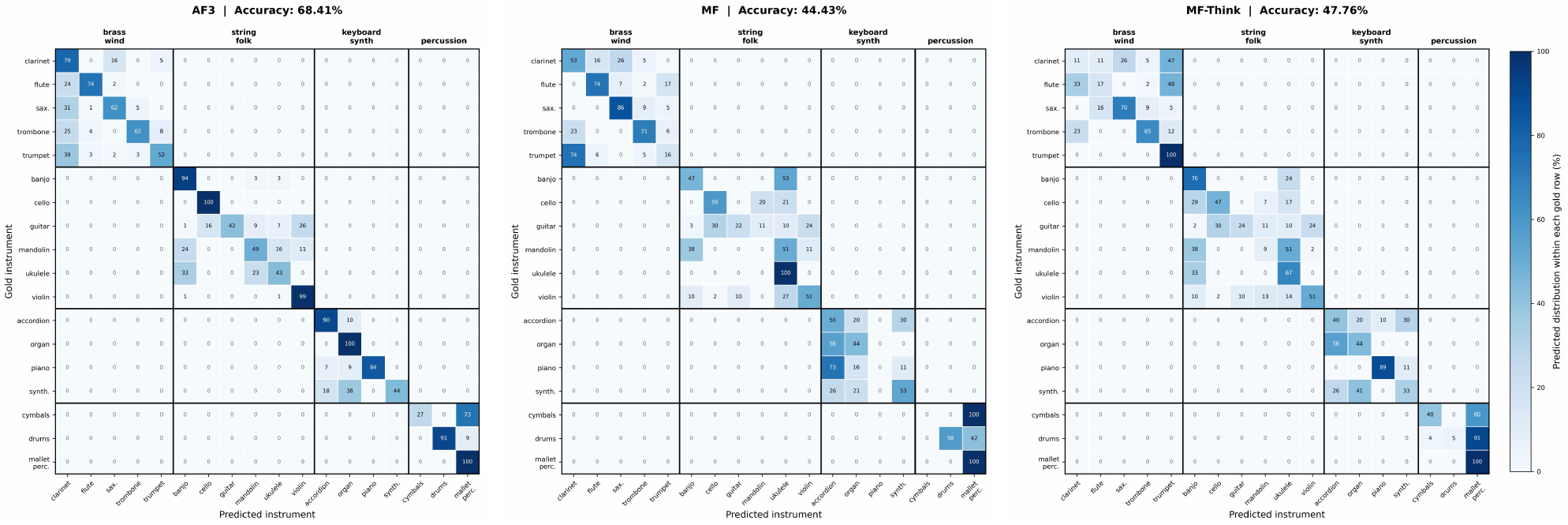}
\caption{
Row-normalized instrument confusion matrices for MF, MF-Think, and AF3 on the confusion-aware instrument discrimination benchmark.
Rows denote gold instruments and columns denote predicted instruments.
Boundary lines (black) indicate predefined confusable instrument groups.
}
\label{fig:confusion_group_boundary_1051}
\vspace{0.2cm}
\end{figure*}

In this section, we analyze model behavior beyond aggregate accuracy, focusing on response imbalance and structured confusions.
The diagnostic tasks introduced above are designed not only to change difficulty, but also to expose failure modes and confusion patterns that binary instrument-presence QA does not reveal.

\subsection{Response Bias Beyond Accuracy}
\label{sec:bias_metric}

To quantify response imbalance, we use a simple max--min prediction-rate gap.
Given a set of possible responses and the model's prediction rate $p_i$ for each response $i$, the bias score is defined as:
$
    \mathrm{Bias} = \max_i p_i - \min_i p_i.
$
For the multiple-choice benchmark, responses correspond to the first and second candidate positions.
For temporal localization, responses correspond to the three time ranges.
A larger value indicates that the model disproportionately selects a subset of available responses, even when the benchmark is balanced by construction~\cite{zang2025you}.

\subsection{Option-Position Bias in Multiple Choice}
\label{sec:mc_bias}

The confusion-aware instrument discrimination benchmark reveals performance gaps hidden by binary QA.
Among models evaluated on binary QA, all achieve high accuracy; however, the Flamingo-family models degrade substantially when asked to distinguish confusable instruments: MF reaches 44.43\%, MF-Think 47.76\%, and AF3 68.41\%.

\begin{table}[t!]
\centering
\caption{Response-bias analysis on the confusion-aware instrument discrimination benchmark. Option rates report how often the model selects the first or second candidate. Position gap is the absolute difference between the two rates.}
\label{tab:mc_bias}
\resizebox{\linewidth}{!}{
\begin{tabular}{lcccc}
\toprule
Model & Acc. & Option 1 & Option 2 & Pos. Gap \\
\midrule
MF & 44.43 & 55.47 & 44.53 & 10.94 \\
MF-Think & 47.76 & 68.13 & 31.87 & 36.25 \\
Qwen2.5-Omni & 86.11 & 53.95 & 46.05 & 7.90 \\
AF3 & 68.41 & 68.41 & 31.59 & 36.82 \\
GPT-4o-audio & 87.73 & 49.86 & 50.14 & 0.29 \\
Gemini 2.5 Pro & 83.82 & 48.85 & 51.15 & 2.31 \\
Gemini 2.5 Flash & 83.92 & 47.38 & 52.62 & 5.23 \\
\bottomrule
\end{tabular}
}
\end{table}

Table~\ref{tab:mc_bias} shows that the performance degradation of Flamingo-family models is accompanied by option-position bias, a known robustness issue in multiple-choice evaluation~\cite{zheng2024large,pezeshkpour2024large}.
MF-Think and AF3 strongly prefer the first option, with position gaps above 36 pp, while MF shows a smaller but visible gap.
In contrast, GPT-4o-audio~\cite{hurst2024gpt}, Gemini models~\cite{comanici2025gemini}, and Qwen2.5-Omni maintain more balanced option distributions.
Moving from yes/no QA to two-way discrimination therefore exposes both weaker instrument discrimination and sensitivity to candidate order.
Appendix~\ref{app:prompt_variation} provides an additional MF analysis under alternative answer interfaces and candidate orders, where the model exhibits a strong first-option preference.

\subsection{Instrument Confusion Structure}
\label{sec:instrument_confusion}

Figure~\ref{fig:confusion_group_boundary_1051} shows row-normalized confusion matrices for the three lowest-performing models on the discrimination benchmark: MF, MF-Think, and AF3.
Errors are not uniformly distributed across instruments; each model exhibits structured confusion patterns within the predefined groups.

MF and MF-Think show clear instrument-label preferences; MF over-selects ukulele and accordion, while MF-Think over-selects trumpet, ukulele, and mallet percussion.
AF3 performs better overall, but still shows uneven group-specific errors, such as over-prediction of clarinet in the brass/wind group.

These patterns show that multiple-choice failures cannot be explained by option-position bias alone~\cite{naik2018stress}, but also reflect label-level preferences and group-specific confusions.

\subsection{Temporal Response Bias}
\label{sec:temporal_bias}

Table~\ref{tab:temporal_bias} reports temporal localization accuracy and prediction rates over the three time ranges.
Because the benchmark is balanced by construction, strong deviations from one-third indicate time-range response bias.

\begin{table}[t]
\centering
\caption{Temporal localization accuracy and prediction distribution on the temporal instrument localization benchmark. Bias is computed as the difference between the maximum and minimum prediction rates across the three time ranges.}
\label{tab:temporal_bias}
\resizebox{\linewidth}{!}{
\begin{tabular}{lccccc}
\toprule
Model & Acc. & 0--10s & 10--20s & 20--30s & Bias \\
\midrule
MF & 52.14 & 53.65 & 10.53 & 35.82 & 43.11 \\
MF-Think & 44.90 & 34.37 & 29.23 & 36.41 & 7.18 \\
Qwen2.5-Omni & 72.90 & 33.78 & 51.41 & 14.81 & 36.60 \\
AF3 & 33.70 & 1.15 & 1.98 & 96.87 & 95.73 \\
GPT-4o-audio & 57.50 & 14.53 & 29.00 & 53.14 & 38.61 \\
Gemini 2.5 Flash & 70.30 & 29.90 & 39.93 & 27.66 & 12.27 \\
Gemini 2.5 Pro & 86.28 & 30.40 & 33.81 & 34.03 & 3.63 \\
\bottomrule
\end{tabular}
}
\end{table}

The results reveal distinct temporal failure modes.
AF3 shows the most extreme time-range preference, predicting 20--30 seconds for 96.87\% of examples.
GPT-4o-audio also favors the final segment, while MF over-selects 0--10 seconds and rarely predicts 10--20 seconds. 
Qwen2.5-Omni achieves higher accuracy, but its predictions concentrate on 10--20 seconds and under-represent 20--30 seconds.

The comparison also shows that response bias and temporal grounding are separate axes.
MF-Think has the most balanced prediction distribution, with a bias score of only 7.18, but its accuracy remains low at 44.90\%.
Gemini 2.5 Pro shows the most favorable pattern, combining the highest accuracy with prediction rates closest to the balanced gold distribution.

Overall, the temporal benchmark reveals errors hidden by instrument-presence QA: a model may recognize that an instrument appears somewhere, yet fail to localize when it appears.
Reporting prediction distributions alongside accuracy is therefore important for separating temporal grounding from time-range response bias.

%% file: sections/sec05_conclusion.tex
\section{Conclusion and Future Work}\label{sec:conclusion}

We presented an OpenMIC-derived diagnostic sequence for probing instrument grounding in music audio-language models.
Starting from binary instrument-presence QA, we extended the evaluation to genre-prior-reduced presence QA, confusion-aware instrument discrimination, long-context multi-label recognition, and temporal instrument localization.
Across these settings, we showed that high binary QA accuracy can hide systematic failure modes, including option-position, instrument-label, and temporal-range biases.
These findings suggest that instrument-centered music understanding should be evaluated through multiple diagnostic axes rather than a single aggregate accuracy.
In summary, our results caution against interpreting strong binary-QA performance as sufficient evidence of grounded instrument understanding. 
Future work should extend this diagnostic approach to broader dimensions of music understanding, including harmony, structure, lyrics, and fine-grained temporal reasoning.

%% file: sections/sec10_appendix.tex
\section{Appendix}

\subsection{OpenMIC-2018 Data Format}
\label{app:openmic_format}

OpenMIC-2018 provides 10-second music clips with instrument-level relevance annotations. The aggregated label file used in this work contains 41,534 clip--instrument annotations over 20,000 unique clips. As summarized in Table~\ref{tab:openmic_label_schema}, each row corresponds to one observed clip--instrument pair and includes a clip identifier, an instrument label, an aggregated relevance score, and the number of annotator responses used for aggregation.

\begin{table}[h]
\centering
\caption{Fields in the OpenMIC-2018 aggregated label file.}
\label{tab:openmic_label_schema}
\begin{tabular}{ll}
\toprule
Column & Description \\
\midrule
\texttt{sample\_key} & OpenMIC clip identifier \\
\texttt{instrument} & Annotated instrument class \\
\texttt{relevance} & Aggregated confidence score for the clip--instrument pair \\
\texttt{num\_responses} & Number of annotator responses used for aggregation \\
\bottomrule
\end{tabular}
\end{table}

Table~\ref{tab:openmic_clip_example} shows example annotations for individual clips. A single clip can have multiple observed instrument annotations, each stored as a separate row with its own relevance score.

\begin{table}[h]
\centering
\caption{Example clip--instrument annotations from the OpenMIC-2018 aggregated label file.}
\label{tab:openmic_clip_example}
\begin{tabular}{llcc}
\toprule
\texttt{sample\_key} & \texttt{instrument} & \texttt{relevance} & \texttt{num\_responses} \\
\midrule
000046\_3840 & clarinet & 0.17105 & 3 \\
000046\_3840 & flute & 0.00000 & 3 \\
000046\_3840 & trumpet & 0.00000 & 3 \\
000135\_483840 & saxophone & 0.14705 & 3 \\
000135\_483840 & voice & 1.00000 & 3 \\
000135\_483840 & trumpet & 0.00000 & 3 \\
000182\_145920 & piano & 0.00000 & 3 \\
000182\_145920 & voice & 1.00000 & 3 \\
\bottomrule
\end{tabular}
\end{table}

The instrument vocabulary contains 20 classes: accordion, banjo, bass, cello, clarinet, cymbals, drums, flute, guitar, mallet percussion, mandolin, organ, piano, saxophone, synthesizer, trombone, trumpet, ukulele, violin, and voice.

We use only extreme relevance values as high-confidence labels. A relevance score of 1.0 is treated as a positive label, indicating that the instrument is present, while a relevance score of 0.0 is treated as a negative label, indicating that the instrument is absent. Intermediate relevance values are excluded from gold-label construction to reduce annotation ambiguity. OpenMIC also provides clip-level metadata, including track and genre information, which is used only for the genre-prior-reduced hard set.

\subsection{Confusable Instrument Groups}
\label{app:confusable_groups}

The confusable instrument groups are manually defined during benchmark construction and are not part of the original OpenMIC-2018 annotation file. OpenMIC provides clip--instrument relevance annotations, while the \texttt{group\_name} field is added in our benchmark metadata to support confusion-aware candidate sampling. The manually defined groups are summarized in Table~\ref{tab:confusable_groups}.

These groups are used in the confusion-aware two-choice benchmark and the strict 30-second choose-all benchmark. In both cases, candidates are sampled within the same group to make the alternatives acoustically or musically related. In the two-choice benchmark, one positive and one negative instrument are sampled from the same group. In the strict 30-second benchmark, all four candidate instruments are sampled from the same group, so that the model must distinguish between related candidates rather than choose from unrelated instruments.

These groups should not be interpreted as perceptually validated human-confusion groups. They are intended to create related candidate sets that are more diagnostic than randomly sampled unrelated instruments. Accordingly, the resulting confusion matrices characterize model behavior within our manually defined candidate groups, rather than human perceptual confusability. Future work should validate such groupings through controlled listening studies.

\begin{table}[h]
\centering
\caption{Manually defined confusable instrument groups used for candidate sampling.}
\label{tab:confusable_groups}
\begin{tabular}{ll}
\toprule
Group & Instruments \\
\midrule
\texttt{string\_folk} & banjo, cello, guitar, mandolin, ukulele, violin \\
\texttt{brass\_wind} & clarinet, flute, saxophone, trombone, trumpet \\
\texttt{keyboard\_synth} & accordion, organ, piano, synthesizer \\
\texttt{percussion} & cymbals, drums, mallet percussion \\
\bottomrule
\end{tabular}
\end{table}

\subsection{Binary Instrument-Presence QA Benchmark}
\label{app:binary_qa}

The binary instrument-presence QA benchmark contains 9,332 yes/no question-answer pairs derived from 4,666 OpenMIC-2018 clips. Each clip contributes one positive and one negative question, yielding a balanced set with 4,666 ``Yes'' and 4,666 ``No'' answers. Positive questions use instrument labels with relevance score 1.0, while negative questions use labels with relevance score 0.0.

Each instance consists of an audio clip, a target instrument, and a yes/no question of the form: ``Is there a [instrument] in this audio clip?'' Model outputs are evaluated by exact-match accuracy after answer normalization. The core CSV fields are summarized in Table~\ref{tab:binary_csv_schema}, and representative examples are shown in Table~\ref{tab:binary_examples}.

\begin{table}[h]
\centering
\caption{Core CSV fields for the binary instrument-presence QA benchmark.}
\label{tab:binary_csv_schema}
\begin{tabular}{ll}
\toprule
Column & Description \\
\midrule
qa\_id & Unique identifier of the QA instance \\
sample\_key & OpenMIC-2018 clip identifier \\
audio\_path & Path to the audio file \\
instrument & Target instrument in the question \\
question & Natural-language yes/no question \\
gold\_answer & Ground-truth answer, Yes or No \\
label\_type & Positive or negative label type \\
relevance & OpenMIC relevance score \\
source\_dataset & Source dataset name \\
setting & Benchmark construction setting \\
\bottomrule
\end{tabular}
\end{table}

\begin{table}[h]
\centering
\caption{Representative examples from the binary instrument-presence QA benchmark.}
\label{tab:binary_examples}
\begin{tabular}{lllll}
\toprule
sample\_key & instrument & gold & label type & relevance \\
\midrule
000135\_483840 & voice   & Yes & positive & 1.0 \\
000135\_483840 & trumpet & No  & negative & 0.0 \\
000182\_145920 & voice   & Yes & positive & 1.0 \\
000182\_145920 & piano   & No  & negative & 0.0 \\
\bottomrule
\end{tabular}
\end{table}

This construction reduces answer-prior bias by balancing positive and negative questions at the clip level. Since each selected clip contributes both a positive and a negative query, models cannot achieve high performance by always favoring one answer class.

\subsection{Genre-Prior-Reduced Presence QA}
\label{app:genre_hard}

The genre-prior-reduced hard set contains 590 binary QA examples selected from the main benchmark. It is designed to reduce cases that can be answered using simple genre-instrument associations. We use examples with available genre metadata, split them into 70\% training and 30\% test partitions, and fit a genre-prior baseline on the training split. The baseline estimates the positive-answer rate for each genre-instrument pair, with an instrument-level fallback for unseen pairs. Test examples incorrectly answered by this baseline are retained as hard cases.

Each row follows the binary QA format and additionally stores the metadata used by the genre-prior baseline. The core CSV fields are summarized in Table~\ref{tab:genre_hard_csv_schema}, and representative examples are shown in Table~\ref{tab:genre_hard_examples}. All examples use the same prompt template as the main binary benchmark: ``Is there a [instrument] in this audio clip?''

\begin{table}[h]
\centering
\caption{Core CSV fields for the genre-prior-reduced hard set.}
\label{tab:genre_hard_csv_schema}
\begin{tabular}{ll}
\toprule
Column & Description \\
\midrule
qa\_id & Unique identifier of the QA instance \\
sample\_key & OpenMIC-2018 clip identifier \\
audio\_path & Path to the audio file \\
instrument & Target instrument in the question \\
question & Natural-language yes/no question \\
gold\_answer & Ground-truth answer, Yes or No \\
label\_type & Positive or negative label type \\
relevance & OpenMIC relevance score \\
primary\_genre & Genre used by the prior baseline \\
yes\_rate & Genre-instrument positive rate \\
inst\_yes\_rate & Instrument-level fallback positive rate \\
final\_yes\_rate & Final positive rate used for prediction \\
genre\_prior\_pred & Genre-prior baseline prediction \\
genre\_prior\_correct & Whether the baseline prediction is correct \\
\bottomrule
\end{tabular}
\end{table}

\begin{table}[h]
\centering
\caption{Representative examples from the genre-prior-reduced hard set.}
\label{tab:genre_hard_examples}
\begin{tabular}{llllll}
\toprule
sample\_key & instrument & gold & genre & prior pred. & final rate \\
\midrule
111817\_268800 & cymbals  & Yes & International & No  & 0.333 \\
016747\_334080 & violin   & Yes & Afrobeat      & No  & 0.000 \\
017608\_506880 & cello    & Yes & Asia-Far East & No  & 0.418 \\
057853\_0      & violin   & No  & Pop           & Yes & 0.667 \\
\bottomrule
\end{tabular}
\end{table}

This subset does not remove all possible shortcuts. Instead, it specifically filters out examples solved by a simple genre-instrument prior, thereby increasing the need for audio-grounded instrument recognition.

\subsection{Confusion-Aware Instrument Discrimination}
\label{app:confusion_mc}

The confusion-aware two-choice benchmark contains 1,051 instrument discrimination examples. Each example presents two candidate instruments from a predefined confusable group: one positive instrument with relevance score 1.0 and one negative instrument with relevance score 0.0. The model must output the name of the instrument that is present.

Each instance stores the present instrument, the absent confusable candidate, and the candidate order shown to the model. Although option-position metadata is retained for analysis, evaluation is based on matching the predicted instrument name to the gold instrument name rather than on A/B labels. The core CSV fields are summarized in Table~\ref{tab:confusion_csv_schema}, and representative examples are shown in Table~\ref{tab:confusion_examples}.

All examples use the prompt template: ``Which instrument is present in this audio clip? Candidate instruments: [instrument 1], [instrument 2]. Answer with only one instrument name from the candidates.''

\begin{table}[h]
\centering
\caption{Core CSV fields for the confusion-aware two-choice name-answer benchmark.}
\label{tab:confusion_csv_schema}
\begin{tabular}{ll}
\toprule
Column & Description \\
\midrule
mc\_id & Unique identifier of the instance \\
sample\_key & OpenMIC-2018 clip identifier \\
audio\_path & Path to the audio file \\
group\_name & Confusable instrument group \\
gold\_instrument & Present instrument with relevance score 1.0 \\
negative\_instrument & Absent candidate with relevance score 0.0 \\
options & Candidate instruments shown to the model \\
gold\_answer & Correct instrument name \\
ab\_gold\_answer & Position metadata of the correct candidate \\
label\_type & Name-answer evaluation format \\
\bottomrule
\end{tabular}
\end{table}

\begin{table}[h]
\centering
\caption{Representative examples from the confusion-aware two-choice benchmark.}
\label{tab:confusion_examples}
\begin{tabular}{llll}
\toprule
sample\_key & group & candidates & gold instrument \\
\midrule
000386\_65280 & brass\_wind & clarinet|saxophone & saxophone \\
000739\_0 & string\_folk & guitar|violin & guitar \\
001430\_291840 & keyboard\_synth & organ|synthesizer & synthesizer \\
001378\_34560 & percussion & drums|mallet\_percussion & drums \\
\bottomrule
\end{tabular}
\end{table}

This benchmark removes the yes/no response format and tests whether models can discriminate between acoustically or semantically confusable instruments.

\subsection{Answer-Interface and Candidate-Order Variation}
\label{app:prompt_variation}

We conduct a targeted prompt-variation analysis on MF to examine whether the option-position behavior observed in Section~\ref{sec:mc_bias} persists under different candidate orders and answer interfaces. We use the same confusion-aware two-choice benchmark, keeping the audio clips, candidate instruments, and gold labels fixed. We vary only the displayed candidate order and the answer format.

The main discrimination experiment in Sections~2.3 and~3.2 uses the direct instrument-name format, where the model outputs the name of the candidate instrument that is present. Here, we compare four variants: (i) the original direct instrument-name prompt, (ii) a direct instrument-name prompt with the candidate order swapped, (iii) an A/B answer prompt with the original candidate order, and (iv) an A/B answer prompt with the candidate order swapped. For all variants, model outputs are mapped back to instrument names before evaluation. The results are shown in Table~\ref{tab:prompt_variation_mf}.

\begin{table}[h!]
\centering
\caption{Prompt-variation analysis on MF for the confusion-aware two-choice benchmark. Option rates report how often the model selects the first or second displayed candidate. Gap denotes the absolute difference between the two option rates.}
\label{tab:prompt_variation_mf}
\begin{tabular}{lccccc}
\toprule
Prompt/interface & Acc. & Opt. 1 & Opt. 2 & Unknown & Gap \\
\midrule
Name, original order & 44.43 & 55.47 & 44.53 & 0.00 & 10.94 \\
Name, swapped order  & 42.06 & 58.71 & 41.29 & 0.00 & 17.41 \\
A/B, original order   & 47.86 & 92.58 & 7.42  & 0.00 & 85.16 \\
A/B, swapped order    & 44.81 & 93.24 & 6.76  & 0.00 & 86.49 \\
\bottomrule
\end{tabular}
\end{table}

The original direct instrument-name condition reproduces the MF result reported in Table~\ref{tab:mc_bias}, confirming that the prompt-variation pipeline is comparable to the main experiment. Across all four variants, MF predicts the first displayed candidate more often than the second, indicating that the first-option preference is not limited to a single candidate ordering. The bias becomes especially severe under the A/B answer interface: MF selects the first displayed candidate in more than 92\% of examples under both original and swapped candidate orders. These results show that MF's multiple-choice behavior is sensitive not only to the candidate order, but also to the response interface itself.

\subsection{Long-Context Multi-label Instrument Recognition}
\label{app:strict30}

The long-context multi-label instrument recognition benchmark contains 1,028 multi-label instrument recognition examples. Each example is constructed by concatenating three 10-second OpenMIC-2018 clips into a 30-second input. Four candidate instruments are shown to the model: two positive instruments with relevance score 1.0 and two negative instruments with relevance score 0.0. The model must select all candidate instruments that appear anywhere in the 30-second audio.

Each row stores the three source clip identifiers, their original audio paths, the positive and negative candidate instruments, the full candidate list, and the complete gold answer set. Evaluation is performed using exact-set accuracy, precision, recall, and F1. The core CSV fields are summarized in Table~\ref{tab:strict_csv_schema}, and representative examples are shown in Table~\ref{tab:strict_examples}.

All examples use the prompt template: ``Listen carefully to the 30-second audio clip. Which instruments are present in this audio clip? Candidate instruments: [instrument 1], [instrument 2], [instrument 3], [instrument 4]. Answer with all instrument names from the candidates that are present, separated by commas.''

\begin{table}[h]
\centering
\caption{Core CSV fields for the long-context multi-label benchmark.}
\label{tab:strict_csv_schema}
\begin{tabular}{ll}
\toprule
Column & Description \\
\midrule
concat\_id & Unique identifier of the concatenated instance \\
group\_name & Confusable instrument group \\
source\_sample\_keys & Three OpenMIC clip identifiers \\
positive\_instruments & Present candidate instruments \\
negative\_instruments & Absent candidate instruments \\
options & Four candidate instruments shown to the model \\
gold\_answers & Complete set of correct instrument names \\
num\_source\_clips & Number of source clips, fixed to 3 \\
duration\_sec & Duration of the concatenated audio, fixed to 30 \\
num\_positive & Number of positive instruments, fixed to 2 \\
num\_negative & Number of negative instruments, fixed to 2 \\
gold\_positions & Option positions of the gold instruments \\
audio\_path & Path to the concatenated 30-second audio file \\
\bottomrule
\end{tabular}
\end{table}

\begin{table}[h]
\centering
\caption{Representative examples from the long-context multi-label benchmark.}
\label{tab:strict_examples}
\begin{tabular}{llll}
\toprule
concat\_id & group & candidates & gold instruments \\
\midrule
000000 & string\_folk & mandolin | ukulele | violin | banjo & mandolin | violin \\
000001 & string\_folk & guitar | cello | violin | ukulele & cello | violin \\
000002 & string\_folk & mandolin | ukulele | cello | guitar & cello | guitar \\
000003 & string\_folk & ukulele | violin | guitar | cello & cello | violin \\
\bottomrule
\end{tabular}
\end{table}

This benchmark is stricter than two-choice discrimination because the model must recover the complete set of present instruments. Exact-set accuracy measures full recovery, while precision, recall, and F1 capture partial recognition.

\subsection{Temporal Instrument Localization}
\label{app:temporal}

The temporal localization benchmark contains 3,579 examples constructed from 30-second concatenated audio inputs. Each input consists of three 10-second OpenMIC-2018 clips arranged into non-overlapping temporal segments. For a target instrument, exactly one segment has a high-confidence positive label with relevance score 1.0, while the other two segments have negative labels with relevance score 0.0. The model must choose the time range in which the target instrument appears, rather than only deciding whether the instrument is present somewhere in the audio.

Each row stores an anonymized audio identifier, the path to the 30-second audio file, the target instrument, the gold time range, and the positive segment index. Segment indices 0, 1, and 2 correspond to 0--10 seconds, 10--20 seconds, and 20--30 seconds. This makes the gold answer directly recoverable from the position of the positive source segment in the concatenated input. The core CSV fields are summarized in Table~\ref{tab:temporal_csv_schema}, and representative examples are shown in Table~\ref{tab:temporal_examples}.

All examples use the prompt template: ``Listen to the full audio carefully and identify the time range where the target instrument is heard. Target instrument: [instrument]. Choose exactly one time range: 0--10 seconds, 10--20 seconds, 20--30 seconds. Answer with only the selected time range.''

\begin{table}[h]
\centering
\caption{Core CSV fields for the temporal localization benchmark.}
\label{tab:temporal_csv_schema}
\begin{tabular}{ll}
\toprule
Column & Description \\
\midrule
item\_id & Unique identifier of the temporal instance \\
anon\_audio\_id & Anonymized audio filename \\
audio\_path & Path to the 30-second audio file \\
instrument & Target instrument to localize \\
group\_name & Instrument group of the target instrument \\
question & Natural-language temporal localization prompt \\
gold\_time\_range & Correct time range answer \\
positive\_segment\_index & Segment index containing the target instrument \\
num\_source\_clips & Number of source clips, fixed to 3 \\
duration\_sec & Duration of the concatenated audio, fixed to 30 \\
\bottomrule
\end{tabular}
\end{table}

\begin{table}[h]
\centering
\caption{Representative examples from the temporal localization benchmark.}
\label{tab:temporal_examples}
\begin{tabular}{llll}
\toprule
item\_id & instrument & group & gold time range \\
\midrule
item\_000000 & mandolin & string\_folk & 20--30 seconds \\
item\_000001 & ukulele & string\_folk & 20--30 seconds \\
item\_000002 & accordion & keyboard\_synth & 0--10 seconds \\
item\_003578 & mallet\_percussion & percussion & 20--30 seconds \\
\bottomrule
\end{tabular}
\end{table}

The benchmark is nearly balanced across the three candidate time ranges, with 1,190 examples labeled as 0--10 seconds, 1,195 as 10--20 seconds, and 1,194 as 20--30 seconds.
This balance prevents accuracy from being dominated by a majority time range and allows temporal response bias to be analyzed separately from localization accuracy.